\documentclass[twocolumn]{pasj01}
\usepackage[round]{natbib}
\Received{$\langle$reception date$\rangle$}
\Accepted{$\langle$acception date$\rangle$}
\Published{$\langle$publication date$\rangle$}
\SetRunningHead{Uchiyama et al. }{Luminous quasars do not live in overdense regions of LBGs at $z\!\sim\!4$}

\begin{document}

\title{{L}uminous Quasars Do Not Live in the Most Overdense Regions of Galaxies at $z\!\sim\!4$}
\author{Hisakazu \textsc{Uchiyama} \altaffilmark{1}, Jun \textsc{Toshikawa}\altaffilmark{2}, Nobunari \textsc{Kashikawa}\altaffilmark{1,2}, 
Roderik \textsc{Overzier}\altaffilmark{3}, Yi-Kuan \textsc{Chiang}\altaffilmark{4}, 
Murilo \textsc{Marinello}\altaffilmark{3}, 
Masayuki \textsc{Tanaka}\altaffilmark{2}, Yuu \textsc{Niino}\altaffilmark{2}, Shogo \textsc{Ishikawa}\altaffilmark{1}, Masafusa \textsc{Onoue}\altaffilmark{1},  Kohei \textsc{Ichikawa}\altaffilmark{2, 5, 6}, Masayuki \textsc{Akiyama}\altaffilmark{7}, Jean \textsc{Coupon}\altaffilmark{8}, Yuichi \textsc{Harikane}\altaffilmark{9,10}, 
Masatoshi \textsc{Imanishi}\altaffilmark{1,2}, Tadayuki  \textsc{Kodama}\altaffilmark{1,2}, Yutaka \textsc{Komiyama}\altaffilmark{1,2}, Chien-Hsiu \textsc{Lee}\altaffilmark{11}, Yen-Ting \textsc{Lin}\altaffilmark{12}, Satoshi \textsc{Miyazaki}\altaffilmark{1,2}, Tohru \textsc{Nagao}\altaffilmark{13},  Atsushi J. \textsc{Nishizawa}\altaffilmark{14}, Yoshiaki \textsc{Ono}\altaffilmark{9},  Masami \textsc{Ouchi}\altaffilmark{9,15}, 
Shiang-Yu \textsc{Wang}\altaffilmark{12}}

\altaffiltext{1}{Department of Astronomical Science, SOKENDAI (The Graduate University for Advanced Studies), Mitaka, Tokyo 181-8588, Japan}
\altaffiltext{2}{Optical and Infrared Astronomy Division, National Astronomical Observatory of Japan, Mitaka, Tokyo 181-8588, Japan}
\altaffiltext{3}{Observat\'orio Nacional, Rua Jos\'e Cristino, 77. CEP 20921-400, S\~ao Crist\'ov\~ao, Rio de Janeiro-RJ, Brazil}
\altaffiltext{4}{Department of Physics \& Astronomy, Johns Hopkins University, Bloomberg Center, 3400 N. Charles St, Baltimore, MD 21218, USA}
\altaffiltext{5}{Department of Astronomy, Columbia University, 550 West 120th Street, New York, NY 10027, USA}
\altaffiltext{6}{Department of Physics and Astronomy, University of Texas at San Antonio, One UTSA Circle, San Antonio, TX 78249, USA}
\altaffiltext{7}{Astronomical Institute, Tohoku University, Aramaki, Aoba, Sendai 980-8578, Japan}
\altaffiltext{8}{Department of Astronomy, University of Geneva ch. d'\'Ecogia 16, CH-1290 Versoix, Switzerland} 
\altaffiltext{9}{Institute for Cosmic Ray Research, The University of Tokyo, 5-1-5 Kashiwanoha, Kashiwa, Chiba 277-8582, Japan} 
\altaffiltext{10}{Department of Physics, Graduate School of Science, The University of Tokyo, 7-3-1 Hongo, Bunkyo, Tokyo 113-0033, Japan}
\altaffiltext{11}{Subaru Telescope, National Astronomical Observatory of Japan, 650 N Aohoku Pl, Hilo, HI96720, USA} 
\altaffiltext{12}{Institute of Astronomy and Astrophysics, Academia Sinica, Taipei 10617, Taiwan} 
\altaffiltext{13}{Research Center for Space and Cosmic Evolution, Ehime University, Bunkyo-cho 2-5, Matsuyama 790-8577, Japan}
\altaffiltext{14}{Institute for Advanced Research, Nagoya University, Chikusaku, Nagoya 464-8602, Japan}
\altaffiltext{15}{Kavli Institute for the Physics and Mathematics of the Universe (Kavli IPMU, WPI), The University of Tokyo, 5-1-5 Kashiwanoha, Kashiwa, Chiba 277-8583, Japan} 

\email{hisakazu.uchiyama@nao.ac.jp}


\maketitle

\begin{abstract}
We present the cross-correlation between 151 luminous quasars ($M_{\mathrm{UV}} < -26$) and 179 protocluster candidates at $z\!\sim\!3.8$, 
extracted from the Wide imaging survey ($\sim121~ $deg$^2$) performed with a part of the Hyper Suprime-Cam Subaru Strategic Program (HSC-SSP). 
We find that only two out of  $151$ quasars reside in regions that are more overdense compared to the average field at $>4\sigma$.   
The distributions of the distance between quasars and the nearest protoclusters 
and the significance of the overdensity at the position of quasars are statistically identical to 
 those found for $g$-dropout galaxies, suggesting that quasars tend to reside in almost the same environment as star-forming galaxies 
 at this redshift. 
Using stacking analysis, we find that the average density of $g$-dropout galaxies around quasars is slightly higher than that around $g$-dropout galaxies 
on $1.0-2.5$ pMpc scales, while at $<0.5$ pMpc that around quasars tends to be lower.  
We also find that quasars with higher UV-luminosity or with more massive black holes tend to avoid the most overdense regions, and that the quasar 
near zone sizes are anti-correlated with overdensity.  
These findings are consistent with a scenario in which the luminous quasar at $z\sim4$ resides in structures that are less massive than those 
expected for the progenitors of today's rich clusters of galaxies, and possibly that luminous quasars may be suppressing star formation in their 
close vicinity. 
\end{abstract}

\section{INTRODUCTION}
Quasars are among the most luminous objects in the Universe and their luminosity is powered by accretion onto supermassive black holes (SMBHs) in the center of their host galaxies.  
 The descendants of the most luminous quasars which have higher black hole mass likely reside in massive dark halos \citep[]{Shen07} and 
host galaxies today according to the M-$\sigma$ relation  \citep[]{Magorrian98, Marconi03}. 
At high redshift, the reservoir of gas feeding the quasars can be supplied by cold streams \citep[e.g.][]{Dekel06, Keres05, Ocvirk08} or major (wet) mergers \citep[e.g.][]{Lin08}. This suggests that the activity of quasars and therefore their SMBH growth may depend not only on their intrinsic properties but also on the surrounding environment beyond the scale of a galaxy. However, it is not known how exactly the large scale environment affects the mass accretion  onto the SMBH whose scale is much smaller than that of galaxy. 

It is generally assumed that the most luminous quasars and galaxies are hosted by massive halos \citep[]{Springel05}. 
Therefore, if quasar activity is triggered by frequent mergers or cold streams, 
it should preferentially occur in the peaks of the matter density field which are rare \citep{Hopkins08}.  
Accordingly, quasars have been thought to act as signposts for high redshift protoclusters (or clusters in the formation) which are thought to evolve into galaxy clusters seen in the local Universe.  
For example, \citet{Husband13} found that three luminous quasars at $z\!\sim\!5$ reside in overdense regions of Lyman break galaxies (LBGs). 
\citet{Morselli14} found more $i$-dropout galaxies in four $z\!\sim\!6$ quasar fields than expected in a blank field. 
\citet{Adams15} reported that about $10$ \% of a sample of luminous quasars at $z\sim4$ reside in overdense regions of Ly$\alpha$ emitters (LAEs). 
\citet{Banados13} and \citet{Mazzucchelli17} used deep narrow- and broad- band imaging to study the environment of $z\!\sim\!5.7$ quasars, and found no enhancement of LAEs compared with average fields, implying either high-$z$ quasars may not trace the most massive dark matter halos, or the strong radiation from the quasar may be suppressing the star formation in galaxies in its immediate vicinity \citep[][and see \S4.2]{Kashikawa07, Bruns12}. 
\citet{Kim09} concluded that only two fields were overdense among five quasar fields at $z\!\sim\!6$. 
\citet{Trainor12} showed that hyper luminous quasars reside in galaxy overdense regions, but the environment around them is not very different from that around much less luminous quasars.  
\citet{Angulo12} show that the $z=0$ descendant halo masses of $z\!\sim\!6$ quasars widely span from cluster- to group-scale masses even if the quasars reside in the most massive objects at that time based on a cosmological $N$-body simulation. 
Recently, \citet{Kikuta17} investigated quasar fields at $z\!\sim\!4.9$ to suggest that high-z quasars are not associated with extreme galaxy overdense regions.  From the above, it is clear that a consistent picture of the environment of high-$z$ quasars still needs to be derived \citep[see][for a review]{Overzier2016}.

The underlying dark halo masses of quasar hosts have been statistically estimated by clustering analysis. 
The optically selected quasar halo masses at $z\!\sim\!1-3$ are consistent with the typical halo mass of  $\!\sim\! 10^{12}~ M_{\odot}$ at that epoch \citep[]{Adelberger05, Coil07, White12}, suggesting they do not reside in very biased regions. 
In fact, recently, \citet{Cai16} found that the masses of protoclusters which are traced by Ly$\alpha$ forest absorption are much more massive than those of typical quasars at $z\!\sim\!2-3$.  
At higher redshift $z>3$, \citet{Shen07} showed that {optical} quasar halo masses are more massive with $M \!\sim\! (4 - 6) \times10^{12}~ h^{-1} M_{\odot}$, implying that almost all luminous quasars could be associated with cluster progenitors.  
On the other hand, \citet{Fanidakis13} used GALFORM \citep[]{Cole00, Fanidakis12, Lacey15}, a semi-analytic model  
that takes into account active galactic nuclei (AGN) feedback suppressing gas cooling in massive halos, to conclude 
 that typical quasars reside in halos whose average masses are $\!\sim\! 10^{11.5}-10^{12}~ h^{-1} M_{\odot}$ at $z>3$;  therefore they are not typically hosted by the most massive halos or the progenitors of the most massive local clusters. 
According to the GALFORM, the quasar halo masses are regulated by AGN feedback; the most massive halos are in the state in which SMBH mass and spin are higher and the accretion rate is lower due to radio mode feedback compared to the less massive halos. As a result, it is difficult for quasars to appear in the most massive halos. 
Recently, \citet{Oogi16} used $\nu$GC, which is another semi-analytic model of galaxy and quasar formation \citep[]{Enoki03, Enoki14, Nagashima05, Shirakata15}, to estimate median quasar halo masses of a few $10^{11}~ M_{\odot}$ at $z=4$.  
These predicted average quasar halo masses are much smaller than expected for progenitors of local clusters, 
and as a consequence, they imply that quasars should not typically be associated with galaxy overdensities at high redshift. 
However, until now, good empirical data on high-$z$ quasar environments, especially, the relation between quasar activity and overdense regions has been lacking.  

\begin{table*} [htb]
\caption{\label{t1} Details of the five fields included in the HSC-SSP Data Release S16A used in this analysis. }
\begin{center}
\begin{tabular}{l|l|l|l} \hline
Name         & range in R.A. (J2000)                                                    & range in Decl. (J2000)                       & effective area [ deg$^2$]    \\ \hline \hline
W-XMMLSS          &  $1^{h}36^{m}00^{s}$ to $3^{h}00^{m}00^{s}$  &     $-6^{\circ}00'00''$ to $-2^{\circ}00'00''$    &  31.26  \\
W-Wide12H  &  $11^{h}40^{m}00^{s}$ to $12^{h}20^{m}00^{s} $  & $-2^{\circ}00'00'' $  to  $2^{\circ}00'00''$   &  17.00  \\
W-GAMA15H &$14^{h}00^{m}00^{s}$ to $15^{h}00^{m}00^{s} $ & $ -2^{\circ}00'00'' $ to $ 2^{\circ}00'00'' $     & 39.27         \\
W-HECTMAP& $15^{h}00^{m}00^{s}$ to $17^{h}00^{m}00^{s} $ & $42^{\circ}00'00'' $ to  $45^{\circ}00'00''$  & 12.60          \\
W-VVDS    & $22^{h}00^{m}00^{s}$ to $23^{h}20^{m}00^{s} $ & $-2^{\circ}00'00'' $ to $ 3^{\circ}00'00'' $  & 20.73            \\ \hline
\end{tabular}
\end{center}
\end{table*}


Until recently, the number of known protoclusters at $z>3$ has been small, $\!\sim\!10$ \citep[]{Chiang13}. 
This means it has been difficult to systematically examine the relation between quasars and their environments. 
A direct technique to detect nearby clusters by looking for the thermal X-ray emission from the intra-cluster medium \citep[][]{Boh01} 
is insensitive at $z>2$, where clusters are expected to be in the growing phase. 
In order to overcome this problem, we have utilized Hyper Suprime-Cam \citep[HSC; ][]{Miyazaki12}, which is an unprecedented wide-field imaging instrument mounted on the 8.2 m Subaru telescope, to build the largest sample of $z\sim4$ protoclusters to date \citep[]{Toshikawa17}. 
The new protocluster sample was derived from data from the Strategic Program for the Subaru Telescope (HSC-SSP : P.I. Miyazaki) currently on-going (Tanaka et al. 2017, submitted).  
Our bias-free wide-field protocluster survey allows us to also investigate the general relationship between quasars and protoclusters, 
which will help to understand the early interplay between structure formation and galaxy and SMBH co-evolution.  
In this paper, we used the 1st release data (DR1) of HSC-SSP of $\!\sim\!121$ square degrees to statistically characterize the environment of 
a large number of optically selected quasars at $z\!\sim\!3.8$ from Sloan Digital Sky Survey (SDSS), by measuring their typical environmental density and the cross-correlation with our protocluster catalog at the same redshift. 

The paper is organized as follows. 
In \S 2, we describe the protocluster sample in the HSC-Wide layer, and our SDSS quasar sample. 
In \S 3, we investigate the correlation between protoclusters and quasars, and discuss whether quasars are statistically good indicators of 
protoclusters or not. We further examine the average radial density profile around quasars on the small scale using a stacking analysis, 
and its dependence on quasar luminosity and black hole mass. 
The implications of our results are discussed in \S 4. Finally, in \S 5 we conclude and summarize our findings. 
We assume the following cosmological parameters: 
$\Omega_{M} = 0.3 $, $\Omega_{\Lambda} = 0.7$, $H_{0} = 70~ $km s$^{-1}$ Mpc$^{-1}$, and magnitudes are given in the AB system.

\section{DATA AND SAMPLE SELECTION}
\subsection{The Subaru HSC-SSP survey}

The Subaru HSC-SSP survey 
started in early 2014, and will spend 300 nights until completion by 2019 or 2020. 
HSC is equipped with 116 2K $\times$ 4K Hamamatsu fully-depleted CCDs, of which 104 CCDs are used to obtain science data over the field-of-view of 1.$^\circ$5 diameter. 
 The present paper is based on the so-called Wide layer of the HSC-SSP, which has wide-area coverage 
 (in the future, the survey area will reach $\sim 1400$  deg$^2$) and high sensitivity through five optical ($g$, $r$, $i$, $z$, $y$) bands. 
The total exposure times range from $10$ min in the $g$, $r$ bands to $20$ min in the $i$, $z$, and $y$ bands. 
The expected $5\sigma$ limiting magnitudes for point sources are ($g$, $r$, $i$, $z$, $y$) = ($26.5$, $26.1$, $25.9$, $25.1$, $24.4$) mag measured in $2.0~ $ arcsec apertures. 
We expect to obtain $>1400$ protocluster candidates at $z\!\sim\!4$ by the end of the survey \citep[]{Toshikawa17, Toshikawa16}.  
The HSC data \citep[DR S16A][]{Aihara17a}, which includes data taken before April 2016, has already produced an extremely wide field image of $>200$ deg$^2$ with a median seeing of $0.\arcsec6 - 0.\arcsec8$ in the Wide layer. The survey design is shown in \citet{Aihara17b}.  
The filter information is given in \citet{Kawanomoto17}.  
Data reduction was performed with the dedicated pipeline hscPipe \citep[version 4.0.2, ][]{Bosch17}, a modified version 
of the Large Synoptic Survey Telescope software stack \citep[][]{Ivezic08, Axelrod10, Juric15}. 
The astrometric and photometric calibrations are associated with the Pan-STARRS1 system \citep[][]{Schlafly12, Tonry12, Magnier13}. 
We use the cModel magnitude, which is measured by fitting two-component, PSF-convolved galaxy models 
 (de Vaucouleurs and exponential) to the source profile \citep[]{Abazajian04}. We measure fluxes and colors of sources with cModel.

\subsection{HSC protocluster sample}
Our new large survey for high-redshift protoclusters based on the unprecedented imaging data produced by the HSC-SSP survey is ongoing. 
We use the catalog of protocluster candidates at $z\!\sim\!4$ as described in the forthcoming paper, \citet{Toshikawa17}. 
We briefly review the key steps of the construction of the protocluster catalog here. 

\begin{figure}
   \begin{center}
      \FigureFile(80mm,50mm){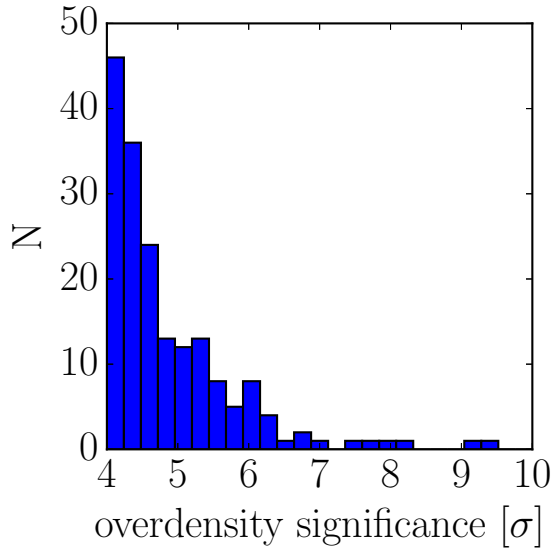}
   \end{center}
   \caption{Histogram of overdensity significance of the 179 protocluster candidates from \citet{Toshikawa17}. The horizontal axis shows the  significance of the overdensities of $g$-dropout galaxies measured within 1.8 arcmin radius circular apertures across the HSC-SSP Wide layer. The maximum significance is $9.37$.}\label{hist_overdensity}
\end{figure}

At first, the $g$-dropout galaxies are selected from 5 independent fields in HSC-SSP S16A DR (i.e. 
W-XMMLSS, W-Wide12H, W-GAMA15H, W-HECTMAP, and W-VVDS ; see Table \ref{t1}) , 
using the following LBG criteria as given in 
\citet{vanderBurg10} : 

\begin{eqnarray}
  1.0  & < & g-r,  \label{eq1}\\ 
          r-i &  < &1.0, \\
1.5 (r-i) & < &(g-r) -0.80, \\
 r  & \le & m_{\mathrm{lim}, 3\sigma}, \\ 
  i  & \le & m_{\mathrm{lim}, 5\sigma},  \label{eq5}
\end{eqnarray}
where $m_{\mathrm{lim}, 3\sigma}$ and $m_{\mathrm{lim}, 5\sigma}$ are $3\sigma$ and $5\sigma$ limiting magnitude, respectively. 
If the objects were not detected in the $g$-band filter at 3$\sigma$, their $g$-band magnitudes were replaced by the corresponding 3$\sigma$ limiting magnitudes. 
We use only homogeneous depth regions, where the limiting magnitudes reach those expected in all $g$-, $r$-, and $i$-bands. 
We carefully mask the regions around bright objects. 
The effective areas are summarized in Table \ref{t1}.  The total effective area achieved was $121$ square  degrees. 
In this area, 259,755 $g$-dropouts down to $i<25.0$ were obtained. This number count is consistent with that of \citet{vanderBurg10}.  
The redshift range of $g$-dropout galaxies is $z\!\sim\!3.8$ and its distribution is expected to be the almost same as that of \citet{vanderBurg10} based on the same color criteria. 

Next, the fixed aperture method was applied to determine the surface density contour maps of $g$-dropout galaxies. 
Apertures with radius of  $1.8$ arcmin, which corresponds to $0.75~ $physical Mpc (pMpc) at $ z \!\sim\! 3.8$, were distributed in the sky of HSC-Wide layer. 
This aperture size is comparable with the typical protocluster size at this epoch with a descendant halo mass of $\gtrsim 10^{14}~ M_{\odot}$ at $z=0$ \citep[]{Chiang13}. 
To determine the overdensity significance quantitatively, we estimated the local surface number density by counting galaxies within the fixed aperture.
The overdensity significance is defined by ($N$-$\bar{N}$)/$\sigma$, where $N$ is the number of the $g$-dropout galaxies in an aperture, 
and $\bar{N}$ and $\sigma$ are average and standard deviation of $N$, respectively.  
Red galaxies at intermediate redshifts and dwarf stars could satisfy our color selection criteria. 
The fraction of contamination from these objects in the $g$-dropout color selection is evaluated to be 25\% at most 
at $i<25.0$ \citep[]{Ono17},  20\% of which originates in photometric error \citep[]{Toshikawa16}. 
The overdensity significance should not be largely affected by the contamination. 



The final protocluster candidates were selected from those regions that had overdensities with $>4\sigma$ significance. 
Although there is a large scatter due to projection effects, surface overdensity significance is closely correlated with descendant halo mass at $z = 0$. 
\citet{Toshikawa16} showed that $ > 76 \%$ of $> 4\sigma$ overdense regions of $g$-dropouts are expected to evolve into dark matter halos with masses of $> 10^{14}~ M_{\odot} $at $z=0$ although there is no clear difference between $< 4\sigma$ overdense regions and fields due to a large scatter.  
It should be noted that the success rate of this technique has already been established by our previous study of the $\!\sim\! 4$  deg$^2$ of the CFHTLS Deep Fields, as the precursor of this HSC protocluster search, followed by Keck/DEIMOS and Subaru/FOCAS spectroscopy \citep[]{Toshikawa16}. 
We carefully checked each $>4\sigma$ overdense region, and removed $22$ fake detections of mainly spiral arms of local galaxies. 
As result, we find $179$ protocluster candidates at $z \!\sim\! 3.8$ in the HSC-Wide layer having overdensity significance ranging from $4$ to $10\sigma$.   
Figure \ref{hist_overdensity} shows the histogram of the overdensity significance of the protocluster candidates. 

\subsection{Quasars}
Our quasar sample was extracted from the Sloan Digital Sky Survey (SDSS) quasar catalog of \citet{Paris17} based on SDSS DR12. 
The SDSS-DR12 quasar catalog is the final SDSS-III quasar catalog. 
The catalog contains $297,301$ spectroscopically confirmed quasars over a wide redshift range of
$0.041 < z < 6.440$ in the area covering approximately $10,200$ deg$^2$ of the sky. 
The quasars have been selected from three different observational programs,  
1) the Baryon Oscillation Spectroscopic Survey \citep[BOSS;][]{Dawson13}, 2)  ancillary programs such as the SDSS-IV pilot survey, 
and 3) objects targeted by chance in other programs, such as the luminous galaxy survey. 
For the details of target selection for spectroscopic observation and quasar confirmation, please see  \citet{Ross12}. 
Within the $g$-dropout redshift range of $z=3.3-4.2$ which corresponds to the $g$-dropout selection function with the value over $0.4$ \citep[]{Ono17}, we find $151$ quasars in the effective areas of the HSC-Wide layer 
used in our analysis.  
The $i$-band absolute magnitudes of this sample range from $\sim -29$ to $-26$ (see Figure \ref{photo_Quasar}). 
We use the reduced one-dimensional (1D) spectral data of the quasars which are available through the SDSS Science Archive Server (SAS) 
to estimate the black hole masses for this sample (\S 3.3). 
The spectral resolution is $R\sim1300-2500$ \citep[]{Paris17}. 
If the quasars have a match in the FIRST radio catalog \citep[][]{Becker95}, whose detection limit is $1$ mJy, within $2.0$ arcsec, the "FIRST\_MATCHED" flag in the SDSS DR12 catalog is set to $1$ \citep[]{Paris17}. 
 The sample contains 8 FIRST- detected quasars (FIRST\_MATCHED $= 1$) with strong radio emission of $L_{\nu}$($6.74$GHz) $ > 5.0 \times 10^{32}$ erg s$^{-1}$ Hz$^{-1}$, which is about $10$ times larger than that of a dichotomy between star forming galaxies and radio active AGNs \citep[][]{Mag02, Mauch07}. It should be noted that the radio property of AGN affects the clustering strength \citep[]{Donoso10}. 

\section{RESULTS}
\subsection{Cross-correlation between protoclusters and quasars} 
\begin{figure}
   \begin{center}
      \FigureFile(90mm,60mm){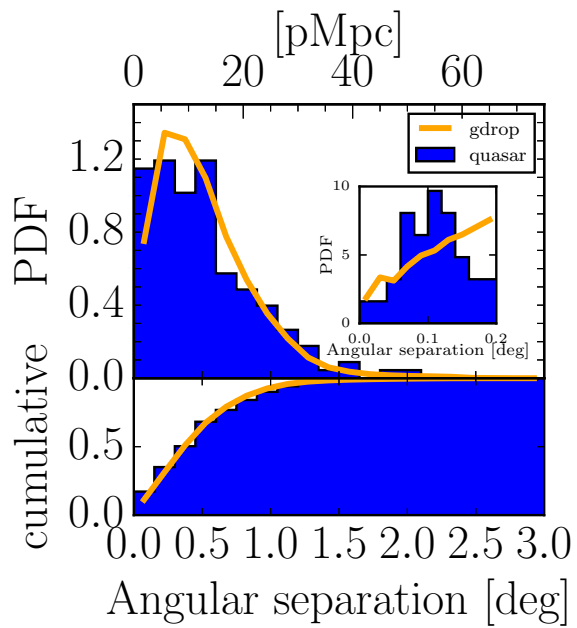}
   \end{center}
   \caption{Histogram of distance from quasars and $g$-dropouts to the nearest protocluster candidates. The blue rods show the case of quasars and the red line shows the case of all $g$-dropout galaxies. On the vertical axis we show the probability distribution function (PDF) and the cumulative PDF. The small chart shows the PDF in the region limited to the small scales, $ < 0.2$ deg. } \label{hist_distance_Quasar}
\end{figure}

First, we measured the projected distance from the quasars to the nearest protoclusters to determine whether the two populations are related.  
The results are shown in Figure  \ref{hist_distance_Quasar}. The blue histogram shows the angular separations between quasars and protoclusters, 
while the orange line shows the separations between $g$-dropout galaxies and protoclusters. 
The $P$-value of the Kolmogorov-Smirnov (KS) test between the two distributions is $0.573$, 
showing that they are not significant different at least on these scales. 
We also focus on the smaller scale $ < 0.2$ degrees to conduct a similar analysis (see small chart in Figure \ref{hist_distance_Quasar}). 
But, the result does not change : the $P$-value in the KS test is $0.100$. 

\begin{figure*}
   \begin{center}
      \FigureFile(180mm,70mm){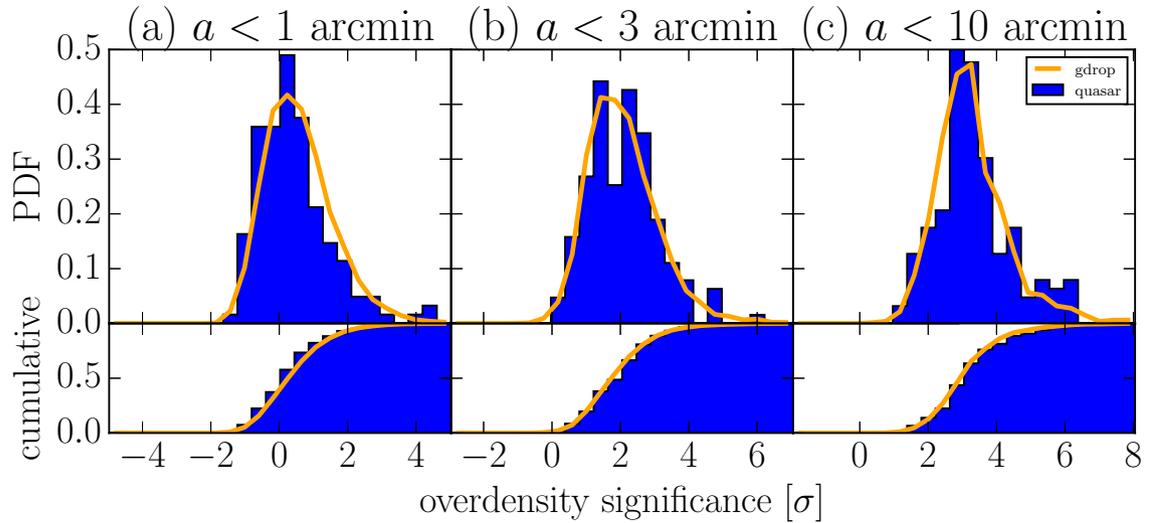}
   \end{center}
   \caption{Histogram of overdensity significance of quasars (blue histograms) and $g$-dropouts (orange lines).  
The panels (a), (b) and (c) show the histograms of the maximum overdensity significance centered on quasars and $g$-dropouts
with the radius of $1$ ($0.42$), $3$ ($1.25$), and $10$ ($4.2$) arcmin (pMpc), respectively. On the vertical axis we show the probability distribution function (PDF) and the cumulative PDF. }\label{hist_overdenisty_Quasar}
\end{figure*}

\begin{table}[htb]
\caption{Statistical Tests of the Correlation of the separations between Quasars and Protoclusters, and $g$-dropouts and Protoclusters shown in Figures 2 and 3. \label{t2}}
\begin{center}
\begin{tabular}{l|l} \hline
                                                         & KS $P$-value \\ \hline \hline 
Figure \ref{hist_distance_Quasar}                                            & 0.573        \\                           
Figure \ref{hist_overdenisty_Quasar} (within $1$ arcmin)             &  $0.105$   \\  
Figure \ref{hist_overdenisty_Quasar} (within $3$ arcmin)            & $0.645$     \\    
Figure \ref{hist_overdenisty_Quasar} (within $10$ arcmin)          & $0.712$     \\ \hline      
\end{tabular}
\end{center}
\end{table}

\begin{figure*}
   \begin{center}
      \FigureFile(160mm,50mm){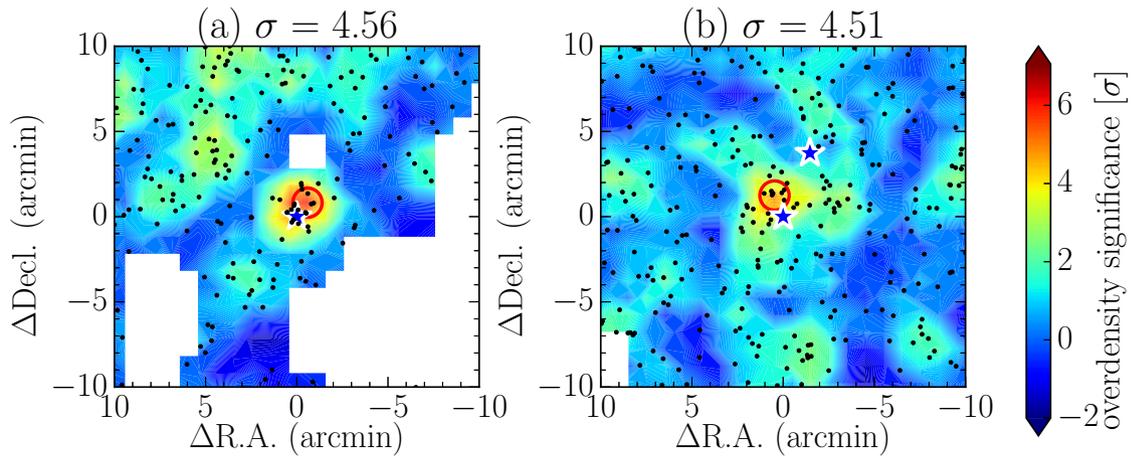}
   \end{center}
   \caption{The two quasars found within 1 arcmin ($0.42$ pMpc) of a protocluster candidate. The quasars are indicated by the blue star symbols. 
The quasar in the center of each panel lies near a $>4\sigma$ maximum overdensity significance region of $\sigma = 4.56$ and $4.51$ in panel (a) and (b), respectively.  The color contours show the overdensity significance map and the red circle shows the local maximum of the region. The black dots show the $g$-dropout galaxies and the mask regions are indicated by the white regions. The size of each panel is $20$ arcmin $\times$ $20$ arcmin. 
The quasar pair at $z\approx3.6$ from \citet{Onoue17} is indicated in the panel (b). }\label{twoQSO}
\end{figure*}

\begin{figure*}
   \begin{center}
      \FigureFile(160mm,50mm){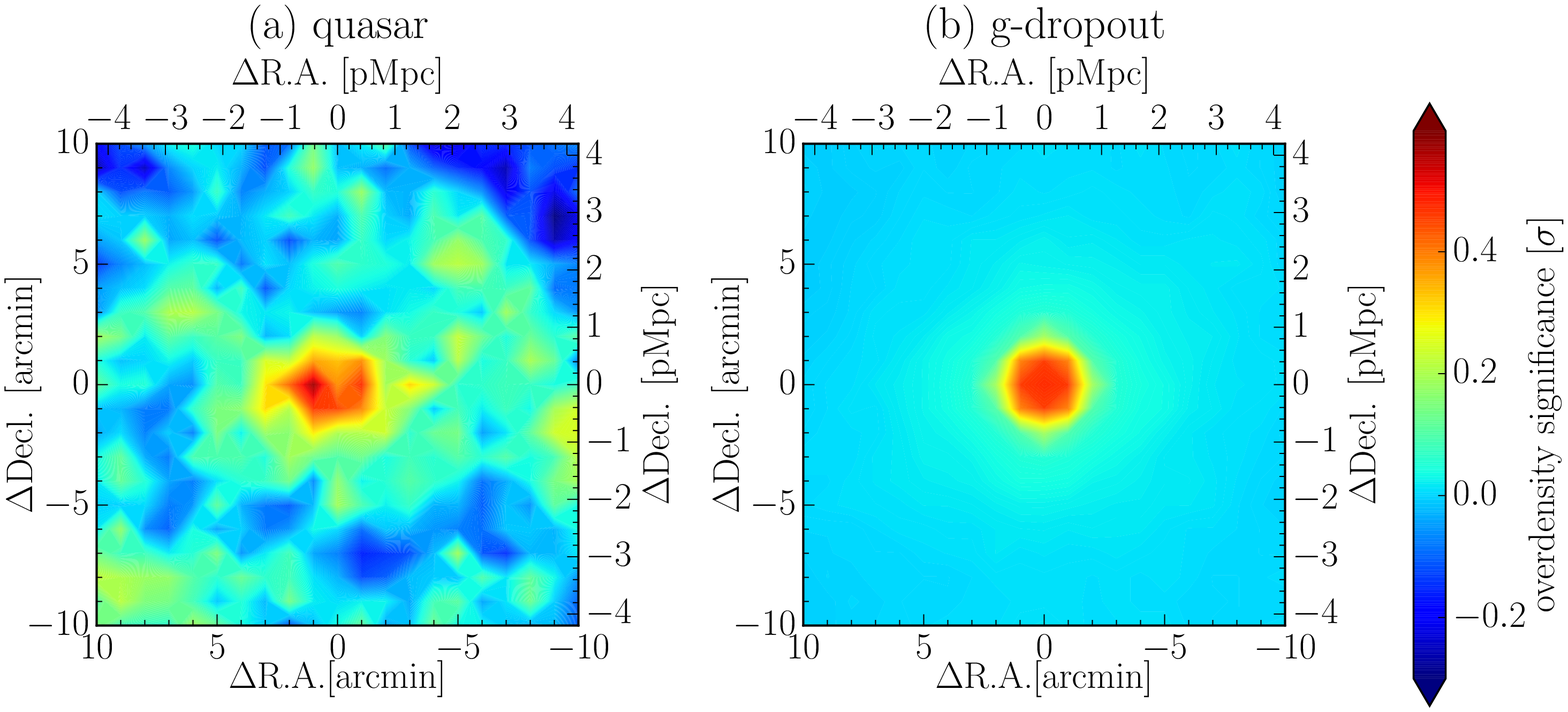}
   \end{center}
   \caption{Median stacking overdensity significance map around quasars/$g$-dropouts. The panel (a) and (b) show the median stacking density map around quasars and $g$-dropouts, respectively. Their objects reside in the center of each panel. The size of each panel is $20$ arcmin $\times$ $20$ arcmin. }\label{stack_map}
\end{figure*}

To measure the environments in detail, 
we also measured the overdensity significances at the exact locations of the quasars. 
Because it is known that quasars and radio galaxies at these redshift are not always right at the center of the overdense regions \citep[e.g.][]{Venemans07}, we measured the maximum overdensity significance within a circle of radius $a = 1$ arcmin ($2.5$ comoving Mpc; cMpc). 
Figure \ref{hist_overdenisty_Quasar} (a) shows the probability distribution of the nearest maximum overdensity significance of quasars and $g$-dropouts. 
We found that only two quasars coincide with protocluster candidates. These protoclusters have overdensity significances of $4.51\sigma$ and $4.56\sigma$, and are shown in Figure \ref{twoQSO}.  It is interestingly to note that one of these protoclusters has two nearby quasars ("quasar pair") at the same redshift of $z\approx3.6$.  This system will be discussed in detail in \citet{Onoue17}.  
The typical radius of protoclusters with $M_{z=0} \gtrsim 10^{14}~ M_{\odot}$ is $\gtrsim 1.8$ arcmin at $z=4$ \citep[]{Chiang13} 
 ; therefore we also check the distributions of the maximum overdensity significance found within a circle of radius $a = 3$ and $a = 10$ arcmin centered on the quasars, shown in panel (b) and panel (c) of Figure \ref{hist_overdenisty_Quasar}, respectively. 
 Again we compare the distribution of overdensity found for quasars with that found for $g$-dropouts. The two distributions are 
 statistically similar at all scales. The results are summarized in Table \ref{t2}.

We expect that the shape of 
the redshift distribution of the quasars is not exactly the same as that of the $g$-dropout galaxies. 
Therefore, the correlations that we are looking for might be diluted by the slight differences in redshift of these samples; however, 
we confirmed that our results did not change even if we extracted $80$ quasars at $z=3.5 - 4.0$ or selected the quasars which satisfy the exact same 
 selection criteria (\ref{eq1}) - (\ref{eq5}) of the $g$-dropouts.  
 This result is robust even if extracting only for the overdense regions with $>5\sigma$ to enhance the purity of the sample: the P-value 
 of KS test between their distributions is 0.437, and focused on the smaller scale $<0.2$ degree, the $P$-value is 0.67. 

\vspace{0.5cm}
\subsection{Stacked densities around quasars}
To further investigate the average environment of quasars,  
we measured the median stacked overdensity significance map around all quasars and $g$-dropout galaxies (Figure \ref{stack_map}). 
Each galaxy overdensity significance map has a $1$ arcmin spatial resolution estimated from the fixed apertures with a radius of $1.8$ arcmin. 
Here, to fairly compare their environments, we count only once those quasars that are also $g$-dropouts.  
The panel (a) and (b) show the overdensity significance map around quasars and $g$-dropouts, respectively. 
 The $g$-dropout galaxies that were used to make these maps are the same as those used to define the overdensity significance. 

\begin{figure}
   \begin{center}
      \FigureFile(90mm,60mm){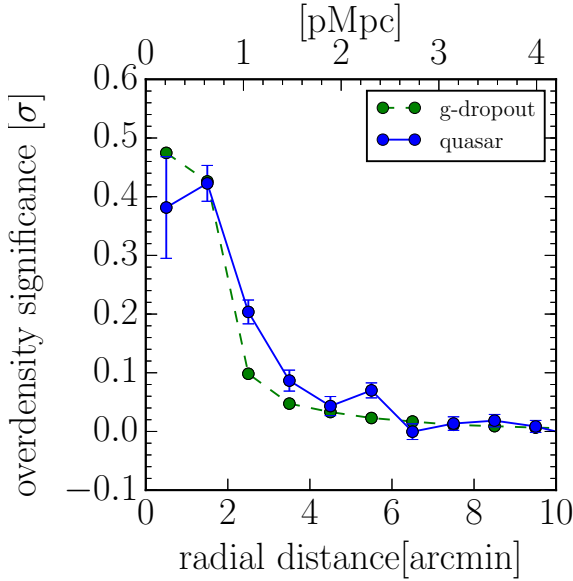}
   \end{center}
   \caption{Radial profile of median stacked overdensity significance around quasars/$g$-dropouts. 
The blue points show the case of quasars. The bars indicate the $1\sigma$ standard error of mean. 
The median of radial distances of $g$-dropouts with 1$\sigma$ standard error bars are shown by green points. }\label{radpro}
\end{figure}

To more clearly show the difference, 
Figure \ref{radpro} shows the radial profile of the median stacked overdensity significance maps of Figure \ref{stack_map}. 
The blue (green) points show the median densities within a circular ring with a width of $1$ arcmin in the case of quasars ($g$-dropouts). 
The error bars show the standard error of the mean. 
 We find an interesting structure : 
the average density around quasars is slightly higher than that of $g$-dropout galaxies in the $1.0-2.5$ pMpc-scale environment at the several sigma level, while on the smaller-scale environment, $<0.5$ pMpc, that of quasars is lower by about one sigma.

\subsection{Rest-UV luminosities and black hole masses of quasars in overdense regions}


\begin{figure*}
   \begin{center}
      \FigureFile(160mm,50mm){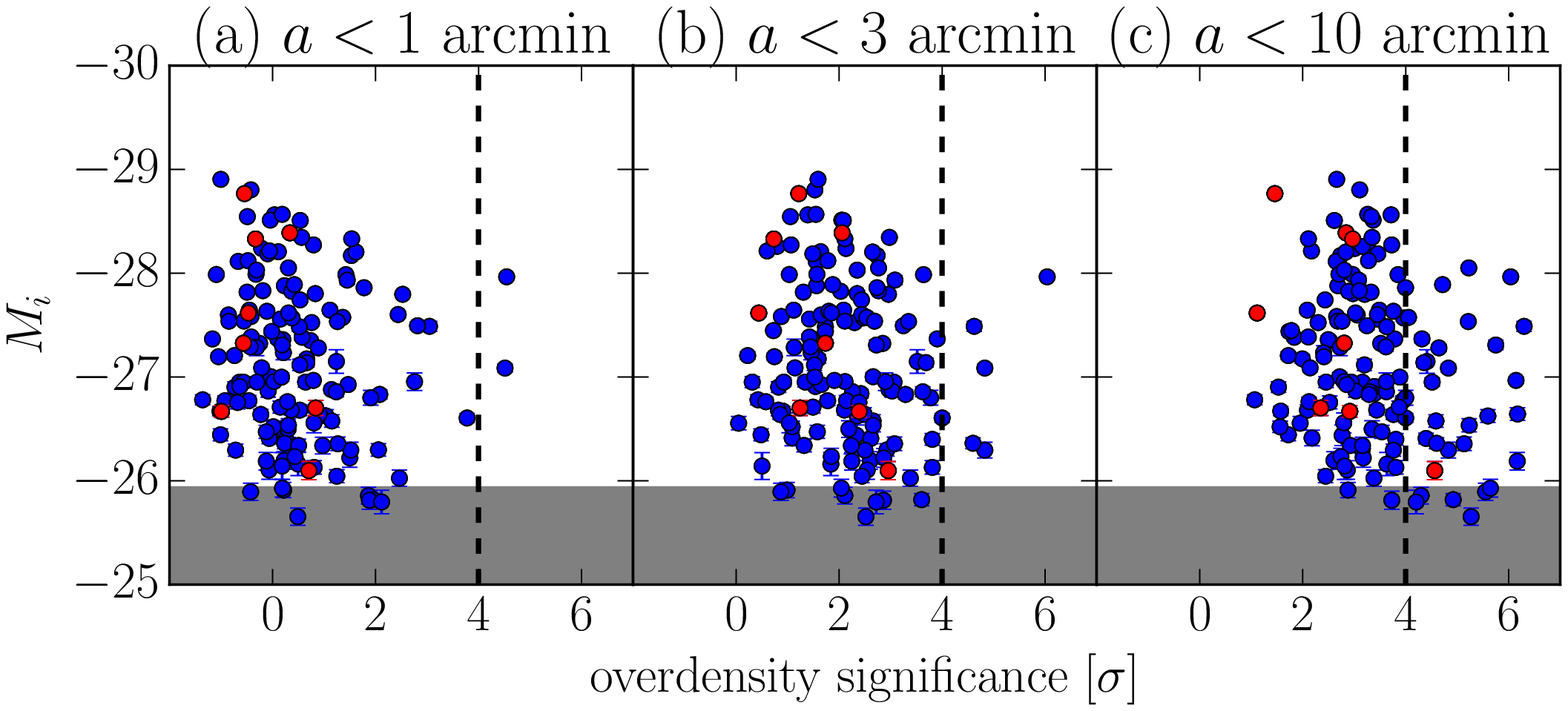}
   \end{center}
   \caption{UV luminosities for overdensity significances about quasars. The panel (a), (b) and (c) show the relation between the $i$-band absolute magnitudes of quasars and the maximum overdensity significances centered on quasars with the radius of $1$, $3$, and $10$ arcmin, respectively. The FIRST detected quasars are also shown by the red points. The $> 4\sigma$ overdense regions are to the right of the dashed line. The gray shade region shows the complete limit, which corresponds to BOSS limiting magnitude of $r<21.8$. }\label{photo_Quasar} 
\end{figure*}


\begin{figure*}
   \begin{center}
      \FigureFile(160mm,50mm){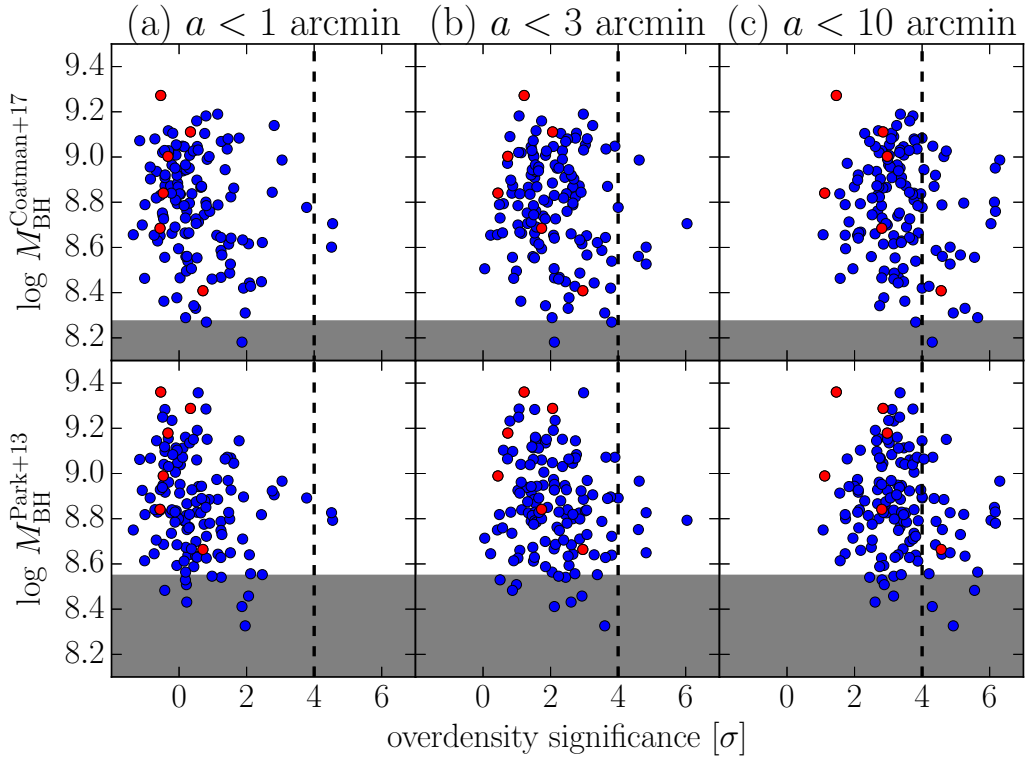}
   \end{center}
   \caption{  Black hole masses for overdensity significances about quasars using two black hole mass estimators, \citet{Coatman17} estimator (first line) and \citet{Park2013} estimator (second line). The panel (a), (b) and (c) show the relation between the CIV-based back hole masses of quasars and the maximum overdensity significances centered on quasars with the radius of $1$, $3$, and $10$ arcmin, respectively.  The FIRST detected quasars are shown by the red points. The $> 4\sigma$ overdense regions are to the right of the dashed line. We shows the complete limit (the gray shade regions), which corresponds to BOSS limiting magnitude of $r<21.8$, assuming that the quasar spectrum energy density $f_{\nu} \propto \nu^{-0.5}$ and $\Delta V_\mathrm{P}(\mathrm{CIV})=4385.06$ and $V_\mathrm{C}^\mathrm{correct}(\mathrm{CIV})=2240.72$ km/s which are the median FWHM of CIV line of SDSS quasar sample at $z=3.3-4.2$.}\label{bh_Quasar}
\end{figure*}

We checked for differences  between the luminosities of quasars inside and outside overdense regions. 
Figure \ref{photo_Quasar} shows the relation between ambient overdense significances and the $i$-band absolute magnitudes of the quasars, 
which correspond to their rest-UV luminosity at this redshift. 
The absolute magnitudes are estimated from the visual inspection redshift of each quasar in the SDSS DR12 catalog, 'Z\_VI'. 
The panels (a), (b) and (c) show the cases of the maximum overdensity significance within the circles of radius of $1$, $3$ and $10$ arcmin centered on  the quasars, respectively. 
In order to show whether our quasar sample is complete, we show the completeness limit (the gray shaded region), which
 corresponds to the BOSS limiting magnitude of $r<21.8$, and assuming that $r-i=0.145$ which is the median color of the SDSS quasar sample at $z=3.3 - 4.2$. 
The Spearman rank correlation test shows correlation coefficients equal to $0.165$, $0.153$, and $0.199$, and the $P$-values equal to  $0.0464$, $0.0611$, and $0.0146$ for the three angular scales, respectively, suggesting weak correlations between quasar luminosity and environment with $\sim2\sigma$ significance. 
These results show that brighter quasars statistically tend to reside in lower density regions of $g$-dropout galaxies. 

 In addition, we evaluate the black hole masses of quasars based on CIV single-epoch measurements \citep[]{Vestergaard2002}. 
For these $z\sim4$ quasars, the best emission lines to estimate the black hole mass do not fall into the SDSS optical window, except for the CIV line. 
It is well-known that the CIV line region is subject to outflows and winds \citep[e.g.][]{Shen08}. 
This is likely to result in large variance and possible biases in the estimated black hole masses.  
Nevertheless, if such non-virial motions occur at random, we can perhaps still statistically address if the environments correlate with the black hole properties. 
We use two black hole mass estimators and compare the results. 
First, we use the calibrated single-epoch black hole estimator \citep[]{Park2013} : 
\begin{eqnarray}
\log \left[\frac{M_{\mathrm{BH}}^{\mathrm{Park+13}}}{M_{\odot}} \right] = \alpha_\mathrm{P} &+& \beta_\mathrm{P} \log \left( \frac{L_{1350}}{10^{44} \mathrm{~erg~ s}^{-1}} \right) \nonumber \\ 
&+& \gamma_\mathrm{P} \log \left( \frac{\Delta V_\mathrm{P}(\mathrm{CIV})}{10^3 \mathrm{~km ~s}^{-1}} \right), 
\end{eqnarray}
where $L_{\mathrm{1350}}=1350\times L_{\lambda}(1350)$ is the monochromatic continuum luminosity at $1350$ \AA $~$ 
and $\Delta V_\mathrm{P}(\mathrm{CIV})$ is the full width at half maximum (FWHM) of CIV which is taken from SDSS DR12 catalog, 'FWHM\_CIV'. 
The parameters $\alpha_\mathrm{P}$, $\beta_\mathrm{P}$, and $\gamma_\mathrm{P}$ are equal to $7.48$, $0.52$, $0.56$, respectively \citep[]{Park2013}. 
Figure \ref{bh_Quasar} shows the relation between the estimated black hole masses of quasars and their $g$-dropout overdensities. 
The Spearman rank correlation test shows correlation coefficients equal to $-0.216$, $-0.138$, and $-0.140$, and $P$-values equal to $0.0136$, $0.111$, and $0.107$ for the three angular scales, respectively. These results hint at a weak anti-correlation between black hole mass and overdensity. 

Recently, \citet{Coatman17} used the quasar spectra which have both Balmer lines and CIV lines to directly compare the velocity widths of their emission lines. They found the following updated estimator :  
\begin{eqnarray}
\log \left[\frac{M_{\mathrm{BH}}^{\mathrm{Coatman+17}}}{M_{\odot}} \right] = \alpha_\mathrm{C} &+& \beta_\mathrm{C} \log \left( \frac{L_{1350}}{10^{44} \mathrm{~erg~ s}^{-1}} \right) \nonumber  \\ 
&+& \gamma_\mathrm{C} \log \left( \frac{\Delta V^{\mathrm{correct}}_\mathrm{C}(\mathrm{CIV})}{10^3 \mathrm{~km ~s}^{-1}} \right), 
\end{eqnarray}
where $\alpha_\mathrm{C}$, $\beta_\mathrm{C}$, and $\gamma_\mathrm{C}$ are equal to $6.71$, $0.53$, $2.00$, respectively.  
Here, the $\Delta V^{\mathrm{correct}}_\mathrm{C}(\mathrm{CIV})$ is the corrected FWHM of the CIV line taking into account the 
blueshift of CIV due to the existence of outflows \citep[]{Coatman17}.  
\citet{Coatman17} found a strong correlation between the blueshift of CIV and the black hole mass ratio estimated from H$\alpha$ and CIV.
Their results show that in order to correct the black hole mass estimated from CIV it is necessary to apply a correction factor based on the blueshift of this line.
The $\Delta V^{\mathrm{correct}}_\mathrm{C}(\mathrm{CIV})$ is obtained from 
\begin{equation}
\Delta V^{\mathrm{correct}}_\mathrm{C}(\mathrm{CIV}) = \frac{\Delta V_\mathrm{C}(\mathrm{CIV})} {(0.41\pm0.03)\times\frac{\mathrm{CIV}_{\mathrm{Blueshift}}}{10^3} + 0.62\pm0.04}
\end{equation}
In order to estimate these parameters we first subtract the continuum shape by fitting a power law ($f_{\lambda} \sim \lambda^{\alpha}$) using the continuum windows of the spectrum at $1445-1455$ \AA$~$ and $1695-1705$ \AA.
The CIV line was fitted using a multi-gaussian model, masking out the regions where absorptions were detected above $1.5\sigma$ as in \citet{Shen08}.
The FWHM and the blueshift (defined as $c \times (1549.48 - \lambda_\mathrm{half})/1549.48$, where $c$ is the speed of light and $\lambda_\mathrm{half}$ is the centroid of CIV estimated from the modeled profile, providing a more accurate measurement. 
For this estimator, the correlation coefficients are equal to $-0.194$, $-0.0969$, and $-0.0985$, and the $P$-values are equal to $0.0273$, $0.263$, and $0.257$, respectively in the Spearman rank correlation test. During this re-analysis we also flagged the most problematic quasars and excluded them from the analysis both for the \citet{Park2013} and \citet{Coatman17} results shown in Figure \ref{bh_Quasar}.   
To summarize, using both estimators we find very similar results.   
The results are the strongest for the smallest scale environment ($< 1$ arcmin) in the sense that 
 quasars with the most massive black holes tend to avoid the most overdense regions. 
These results are summarized in Table \ref{t3} and discussed in next section. 

\begin{table}[htb]
\caption{Spearman rank correlation test results for the relation between absolute $i$-band magnitudes (UV luminosity) and black hole masses of quasars and overdensity within $1$ ($0.42$), $3$ ($1.25$), and $10$ ($4.2$) arcmin (pMpc).  \label{t3}}
\begin{center}
\begin{tabular}{l|l|l} \hline
UV luminosity                                        & $\rho$  \footnotemark[$*$]   &   $P$-value                \\ \hline 
local peak within $1$ arcmin                     & $ 0.165$                       & $ 0.0464  $                        \\
local peak within $3$ arcmin                     & $0.153 $                       &$ 0.0611  $                        \\
local peak within $10$ arcmin                   & $0.199  $                      &$ 0.0146 $                         \\ \hline  \hline 
CIV-based BH mass                                   & $\rho$ \footnotemark[$*$]     &  $P$-value                           \\ \hline 
Coatman et al. (2017)                                &                                         &                    \\ 
local peak within $1$ arcmin                     & $-0.194$                         &  $0.0273  $                        \\
local peak within $3$ arcmin                     & $-0.0969 $                         & $0.263 $                         \\
local peak within $10$ arcmin                   & $-0.0985$                          &$ 0.257$                          \\ \hline  
Park et al. (2013)                                        &                                    &                            \\ 
local peak within $1$ arcmin                     & $-0.216$                         &  $0.0136 $                        \\
local peak within $3$ arcmin                     & $-0.138 $                         & $0.111 $                         \\
local peak within $10$ arcmin                   & $-0.140$                          &$ 0.107$                          \\ \hline 

\multicolumn{3}{@{}l@{}}{\hbox to 0pt{\parbox{85mm}{\footnotesize
\par\noindent
 \footnotemark[$*$] The correlation coefficient in the Spearman correlation test. 
 }\hss}}
\end{tabular}
\end{center}
\end{table}

Finally, we also compare the black hole mass and environment for the FIRST detected quasars.  
The fraction of the FIRST quasars tend to be high in high blackhole mass end as can be seen in \citet{Ichikawa17}, 
and they do not lie in particularly overdense regions, 
although the sample is very small (red points in Figure \ref{photo_Quasar}  and Figure \ref{bh_Quasar}). 

\section{DISCUSSION} 
\subsection{The low coincidence of quasars and protoclusters} 
 We tried to find any positive correlations between quasars and protoclusters at $z\!\sim\!4$
. Only two quasars were found to spatially coincide with protoclusters. 
However, for most quasars there is no correspondence with protoclusters. 
Here we will show how this result can be understood. 

The total comoving cosmic volume covered by our survey ($121$ sq.  degrees between $z=3.3$ and $z=4.2$) is 
about $1.18$ cGpc$^3$. The space density of our 151 quasars which are almost completely observed (see Figure \ref{photo_Quasar} and Figure  \ref{bh_Quasar}) is thus about $3.74 \times10^{-7}~ h^{3}$ cMpc$^{-3}$. 
The total number density of halos capable of hosting such quasars can be evaluated by correcting the quasar number density for duty cycle and viewing angle. 
The net lifetime of a quasar is believed to be about $10^6 - 10^8$ yr from clustering or demographic analysis \citep[e.g.][]{Martini04}. 
The cosmic time between $z=4.2$ and $z=3.3$ is $\sim 0.467$ Gyr. 
So at any given time, we expect that about $0.2 - 20$\% of the halos can host quasars. 
This increases the halo number density to $1.87 \times 10^{-6} - 1.87 \times 10^{-4}~ h^3$ cMpc$^{-3}$. 
Next, we need to correct for viewing angle. 
We expect that at most $50 - 70$ percent of random viewing angles will be along our line-of-sight, producing our luminous quasars \citep[]{Simpson05}. 
Thus the total number density of halos hosting quasar-like host galaxies is $2.67 \times 10^{-6} - 3.74 \times 10^{-4}~ h^3$ cMpc$^{-3}$. 
On the other hand, the number density of local rich clusters of galaxies 
with $>10^{14}~ M_{\odot}$ is about $1.26\times 10^{-5}~ h^3$ cMpc$^{-3}$ \citep[e.g.][]{Rozo09}, 
which is consistent with the halo number density estimated by halo mass functions from \citet{Behroozi13}. 
 The number density of halos capable of hosting quasars at $z\sim4$ is thus close to the number density of clusters today.  

The halo mass of quasars is observationally estimated by clustering analysis as $M_{\mathrm{halo}}\!\sim\! 4 - 6 \times 10^{12}~ h^{-1} M_{\odot}$ at $z>3.5$ \citep[]{Shen07}, which is the minimum halo mass. 
About $30 - 50$ \% of these should evolve into the halos with masses $>10^{14}~ M_{\odot}$ at $z=0$ based on the Extended Press Schechter (EPS) model \citep[]{Press74, Bower91, Bond91, Lacey93}. This estimate is likely to be a lower limit because the mean halo mass is about 3 times as the minimum halo mass \citep[e.g.][]{Ishikawa05}. 
 In principle, the number density and halo mass suggest that our quasars could be associated with protoclusters, 
 assuming that the \citet{Shen07} halo mass estimates are correct. 
However, our results also showed that there is no association between the HSC-SSP Wide protoclusters and the majority of the quasars, 
since only 2 out of 151 quasars  were found to reside near protoclusters. 

Recently,  \citet{Eftek15} derived less massive quasar halo mass, $M_{\mathrm{halo}} = 0.6 \times 10^{12} h^{-1} M_{\odot}$ at $z\sim3$, 
based on the larger sample by BOSS. 
\citet{He17} showed that the average halo mass of $z\!\sim\!4$ less-luminous quasars, $M_{i} \gtsim -26$, detected by HSC 
is $\sim1\times10^{12} h^{-1} M_{\odot}$.  
In the GALFORM \citep[e.g.][]{Lacey15}, the halo mass function of quasars at $z\!\sim\!4$ is predicted to have a peak around  $M_{\mathrm{halo}} \!\sim\! 0.3 - 1.0\times 10^{12}$ $~ h^{-1} M_{\odot}$ \citep[]{Fanidakis13, Orsi16}, which is about 10 times less massive than \citet{Shen07}, with large range of mass distribution. 
\citet{Oogi16} also predicts a few $10^{11}~ M_{\odot}$ for the median halo mass of quasars at $z\!\sim\!4$. 
It should be noted that the clustering strength of quasars is almost independent of their luminosity \citep[e.g. ][]{He17, Shen13, Eftek15}.  
The quasar halo mass of $\!\sim\! 0.3 - 1.0 \times 10^{12}~ h^{-1} M_{\odot}$ at $z\sim4$ provides the probability of only $\sim1-5$\% that the $z=0$ descendant halo mass exceeds $10^{14} M_{\odot}$. 
This result seem to be reasonable for our results, suggesting that a typical quasar halo mass might be $\ltsim 10^{12} h^{-1} M_{\odot}$ at $z\!\sim\!4$. 




If we were to adopt the high halo masses found by \citet{Shen07}, the lack of correspondence between the protoclusters and the quasars is even more striking. 
In this case, it is clear that our $g$-dropout protoclusters must trace the progenitors of much rarer, more massive clusters than the $10^{14}~ M_{\odot}$ halos assumed to be the typical descendant of the \citet{Shen07} quasar hosts. 
For example, if the $g$-dropout protoclusters trace clusters only three times larger than $10^{14}~ M_{\odot}$ would already lead to a 10 times lower number density due to the steep slope of the halo mass function at these high masses. 

\begin{figure}
  \begin{center}
      \FigureFile(80mm,50mm){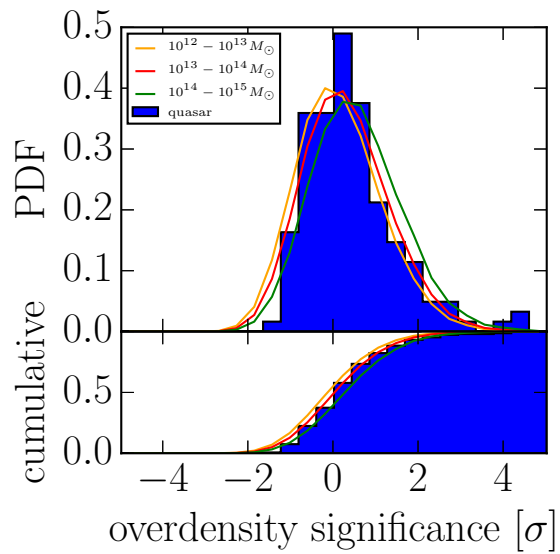}
   \end{center}
  \caption{Correspondence between observational overdensity significances of quasars and their $z=0$ descendant halo masses of quasars and simulated $g$-dropouts. The blue histogram shows the distribution of maximum overdensity significances centered on quasars with the radius of 1 arcmin, which is exactly same as Figure \ref{hist_overdenisty_Quasar} (a). The orange, red, and green lines indicate the simulated $g$-dropout overdensity significance distributions with $z=0$ descendant halo masses $M_{\mathrm{h,z=0}} = 10^{12}~ M_{\odot} - 10^{13}~ M_{\odot}$, $M_{\mathrm{h,z=0}} = 10^{13}~ M_{\odot} - 10^{14}~ M_{\odot}$ and $M_{\mathrm{h,z=0}}>10^{14}~ M_{\odot}$, respectively. Their overdensity significances are the ones simulated for $z\!\sim\!4$. }\label{halomass}
\end{figure}

We can estimate the descendant halo masses at  $z\!\sim\!0$  from the distribution of overdensity significances at the $z\!\sim\!4$ quasar positions (Figure \ref{hist_overdenisty_Quasar} (a)), by using the light cone model  \citep[]{Henriques12}. 
We associate the overdensity significances which simulated $g$-dropout galaxies exist on with their $z=0$ descendant halo masses by tracing back the halo merger tree from $z=0$ to $z \!\sim\! 3.8$. 
Figure \ref{halomass} presents the distribution of overdensity significances of observed quasars at $z\!\sim\!4$ (blue histogram) 
and simulated $g$-dropout galaxies with the descendant halo masses of $M_{\mathrm{h,z=0}} = 10^{12}M_{\odot}  -  10^{13}M_{\odot}$ (orange line), $M_{\mathrm{h,z=0}} = 10^{13}~ M_{\odot}  -  10^{14}~ M_{\odot}$ (red line) and $M_{\mathrm{h,z=0}}>10^{14}~ M_{\odot}$ (green line) in the simulation. 
We found that the predicted descendant halo mass distribution from overdensity significances of our observed quasars at $z\!\sim\!4$ 
are roughly consistent with the mass distribution $M_{\mathrm{h,z=0}} = 10^{13}~ M_{\odot}  -  10^{14}~ M_{\odot}$ with $P$-value $=0.314$, 
while inconsistent with those of $M_{\mathrm{h,z=0}} = 10^{12}~ M_{\odot}  -  10^{13}~ M_{\odot}$ and $M_{\mathrm{h,z=0}}>10^{14}~ M_{\odot}$ with 
$P$-value $=0.00502$ and $0.0332$  in the KS test, respectively. 
Roughly speaking, luminous quasars at $z\!\sim\!4$ resides in the density environments which evolve into halos with the mass of $ \sim 10^{13}~ M_{\odot}  - 10^{14}~ M_{\odot}$.  

The finding that most quasars do not live in the $>4\sigma$ overdense regions is consistent with other results. 
At $z < 2$, many studies on quasar halo mass show that they reside in average halo masses\citep[e.g.][]{Coil07}, implying that they avoid the most overdense regions. This can be understood by well-known environmental relation that passive, red early-type galaxies dominate in overdense regions \citep[]{Baldry04, Dressler80}. 
The brightness of quasars is produced by strong black body radiation (big blue bump) from optically thick accretion disk\citep[e.g.][]{Malkan82}. 
At higher redshift $z\!\sim\! 2 - 3$, protocluster galaxies are on average found to be older and more massive than those in fields 
\citep[]{Steidel05, Koyama13, Cooke14}. 
Thus, quasars may not necessarily appear in mature overdense regions because 
they may lack the wet mergers which lead to quasar activity \citep[]{Lin08}. 



The environment around quasar may not appear as an overdense region due to the strong radiation from quasar \citep[quasar-mode AGN feedback, ][]{Kashikawa07}. This will be discussed in more detail in the next subsection. 

\begin{figure*}
   \begin{center}
      \FigureFile(180mm,70mm){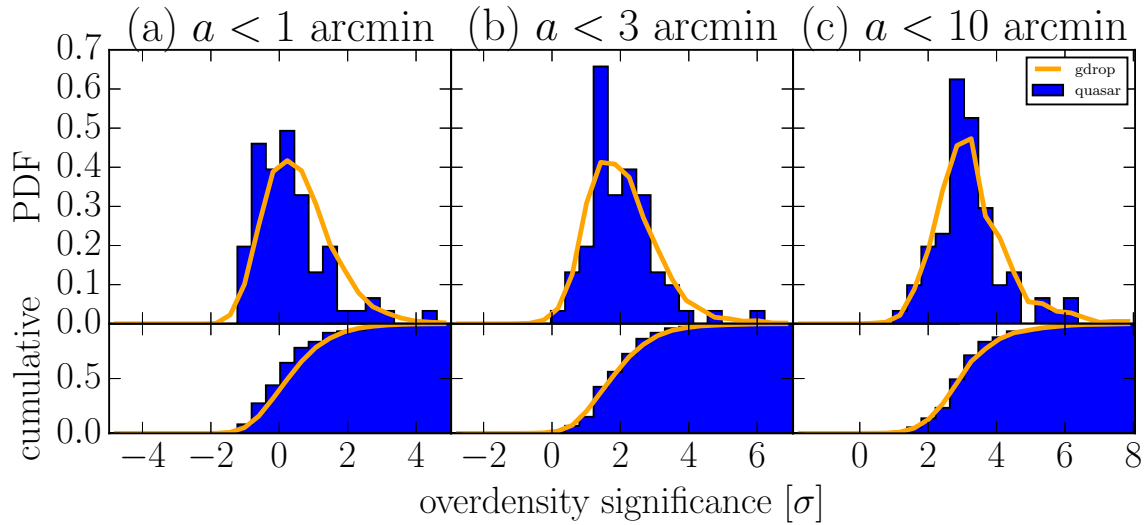}
   \end{center}
   \caption{Histogram of overdensity significance of bright quasars. 
The panel (a), (b) and (c) show the histograms of the maximum overdensity significance centered on quasars 
with the radius of $1$($0.42$), $3$($1.25$), and $10$($4.2$) arcmin(pMpc), respectively.}\label{hist_overdenisty_bright_quasar}
\end{figure*}

\begin{table*}[htb]
\caption{Statistical Test for Correlation of Position between bright quasars and protoclusters\label{t6}}
\begin{center}
\begin{tabular}{l|l|l|l} \hline
                                                             & KS $P$-value                           & Median quasar                & Median $g$-dropout           \\ \hline \hline
local peak within 1 arcmin                     &  0.0392                                     &  0.189                            &  0.474  \\
local peak within 3 arcmin                     & 0.702                                       & 1.78                              &  1.94 \\
local peak within 10 arcmin                   & 0.849                                       & 3.06                              &  3.12 \\ \hline
\end{tabular}
\end{center}
\end{table*}

We can compare our results with \citet{Adams15} who found one overdense region of LBGs ($r < 25.0$) in a sample of $9$ 
luminous SDSS quasars, $M_i < -28.0$, at $z\!\sim\!4$. 
If we choose from our 151 quasars a sample with the same luminosity range as that of their 9 quasars, we find that only
one out of 27 quasars resides in a protocluster. Although the fraction of the quasars in overdensities appears to be three times higher in \citet{Adams15}, 
we note that none of their quasars lie in overdensities as significant as the $>4\sigma$ overdensities that we selected in our work. 
Also, it is interesting to note that the evidence for a physical overdensity in the case of the quasar in \citet{Adams15} only appeared after spectroscopic follow-up, showing that spectroscopic redshifts are crucial to obtain definite answers.    

We also found that no FIRST-detected quasars reside in $>4\sigma$ overdense regions. 
On the other hand, \citet{Hatch14} showed at high statistical significance that radio-loud AGNs reside in denser Mpc-scale environments than similarly massive radio-quiet galaxies at $z=1.3-3.2$. \citet{Venemans07} found that 6 out of 8 radio galaxies reside in overdense regions. 
This difference from our result might be due to not only the above reasons, but also the difference of selection criteria. 
The radio-loud AGN from \citet{Hatch14} and \citet{Venemans07} are about $10$ times more luminous in the radio than 
our FIRST-detected quasars, and the environments where measured in an entirely different manner. 
It is therefore difficult to assess at this point whether our results are consistent or not. 
We are also limited by the small sample size of only 8 quasars. 
In the future, we will investigate in detail the environments of large numbers 
of radio-loud and radio-quiet quasars using the full $1400$ deg$^2$ of the HSC-SSP Wide survey.

\subsection{The relation between density and quasar properties}

\begin{figure}
   \begin{center}
      \FigureFile(80mm,50mm){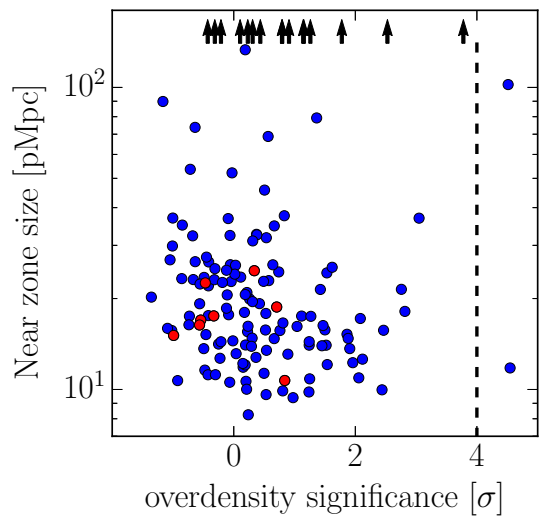}
   \end{center}
   \caption{Near zone size of quasars as a function of the overdensity significance. The red points show the FIRST-detected quasars. The black arrows show quasars with near zone sizes larger than $150$ pMpc. The $> 4\sigma$ overdense regions are to the right of the dashed line.  }\label{nearzone}
\end{figure}

We found that UV brighter quasars statistically tend to reside in lower dense regions of $g$-dropout galaxies. 
In addition, quasars with the most massive black hole tend to avoid overdense regions.  
Figure \ref{hist_overdenisty_bright_quasar} shows the distributions of overdensity significances of bright quasars which are defined by 
brighter than the median of the absolute $i$-band magnitudes of quasar sample. 
In fact, the brighter quasars significantly reside in lower dense regions than $g$-dropout galaxies at least in the $1$ arcmin-scale environment.  
The results are summarized in Table \ref{t6}. 
 Earlier work has suggested that the strong radiation from quasars may provide negative feedback and suppress nearby galaxy formation, 
especially for low-mass galaxies \citep[][]{Kashikawa07}. 
We use the near zone sizes \citep[e.g.][]{Fan06, Mortlock11} of our quasar sample to evaluate the scale of AGN feedback. 
We estimate the intrinsic flux by fitting a single power law to the continuum 
free of line emission at $\sim 1340-1360$ \AA, $1440-1450$ \AA $~$ and $1700-1730$ \AA. 
The IGM transmission curves are calculated by the observed fluxes divided by the intrinsic fluxes 
blueward of Ly$\alpha$, and then fitted by a single power law. 
Then, the near zone size is defined by the proper radius at which the fitted transmission 
first falls below $10$ \% \citep[]{Fan06, Mortlock11}.  
Figure \ref{nearzone} shows the relation between the overdensity significance and near zone size for each quasar. 
Although the scatter is large, it appears that quasars tend to have much smaller near 
zones when their overdensity significance is larger than $\sim 2\sigma$. 
On the other hand, quasars in lower density regions can have the full range of 
near zone sizes. 
The statistical coefficient $\rho$ and the $P$-value are $-0.232$ and $0.00798$ in the Spearman rank correlation test. 
These findings may indicate that perhaps galaxy formation is delayed 
in overdense regions due to the quasar feedback. Another possibility 
is that the near zones in overdense regions are smaller due to the 
much higher gas densities that are more difficult to fully ionize. 

 Considering the viewing angle of type-1 quasars, 
this effect should be the most significant on the number density of galaxies along the quasar line of sights (LOS), 
and become weaker as it goes away from the LOS. 
The tendency towards low-density at $<1$ arcmin as seen on Figures \ref{radpro} and \ref{hist_overdenisty_bright_quasar} 
could also be caused by the effect.   
Our statistical test to see the possible UV luminosity, black hole mass and near zone size dependence on the distance to the nearest overdense region (see Figures \ref{photo_Quasar}, \ref{bh_Quasar}, and \ref{nearzone}) agrees with the hypothesis although projection effects may smear out this small-scale over (under) density signal.

\section{CONCLUSIONS}
We have used the statistically significant sample of $179$ protoclusters from the HSC survey \citep[]{Toshikawa17}.   
 to investigate the spatial correlation between the protoclusters and quasars at $z\!\sim\!4$.  
Note that this sample is around $10$ times as large as the number of all previously known protoclusters. 
We found $151$ SDSS quasars at $z=3.3 - 4.2$ in effective areas of HSC-Wide layer, 
whose redshift range corresponds to $g$-dropout selection function. 
 
We showed, for the first time, that quasars statistically do not reside in the most massive halos at $z\!\sim\!4$ 
, assuming that environment correlates with halo mass. 
We used stacking analysis to find that the average density of $g$-dropout galaxies around quasars is slightly higher than that around $g$-dropout galaxies 
on $1.0-2.5$ pMpc scales, while at $<0.5$ pMpc that around quasars tends to be lower.  
We also find that quasars with higher UV-luminosity or with more massive black holes tend to avoid the most overdense regions, and that the quasar 
near zone sizes are anti-correlated with overdensity.  

We discussed some possibilities for why the most luminous quasars do not live in the most overdense regions : 
(1) wet mergers do not often occur in the mature overdense regions, 
(2) quasar-mode AGN feedback disturb the galaxy formation, 
(3) quasar halo mass at $z\!\sim\!4$ is not massive enough to evolve into cluster halo mass.  

In the future, we will investigate a possible correlation between AGNs and $>1000$ complete protocluster candidates found in HSC-SSP final data release with better statistical accuracy.

\bigskip
This work is based on data collected at the Subaru Telescope and retrieved from the HSC data archive system, which is operated by Subaru Telescope and Astronomy Data Center at National Astronomical Observatory of Japan. 

Alvaro Orsi gives me constructive comments and warm encouragement about AGN modeling in the semi-analytic model. 
Jim Bosch, Hisanori Furusawa, Robert H. Lupton, Sogo Mineo, and Naoki Yasuda give us helpful comments and discussions on the treatment of the HSC data.  NK acknowledges supports from the JSPS grant 15H03645.This work was partially supported by Overseas Travel Fund for Students (2016) of
the Department of Astronomical Science, SOKENDAI (the Graduate University for Advanced Studies).  MT is supported by JSPS KAKENHI Grant Number JP15K17617. Authors RO and YTL received support from CNPq (400738/2014-7). 

The Hyper Suprime-Cam (HSC) collaboration includes the astronomical communities of Japan and Taiwan, and Princeton University. The HSC instrumentation and software were developed by the National Astronomical Observatory of Japan (NAOJ), the Kavli Institute for the Physics and Mathematics of the Universe (Kavli IPMU), the University of Tokyo, the High Energy Accelerator Research Organization (KEK), the Academia Sinica Institute for Astronomy and Astrophysics in Taiwan (ASIAA), and Princeton University. Funding was contributed by the FIRST program from Japanese Cabinet Office, the Ministry of Education, Culture, Sports, Science and Technology (MEXT), the Japan Society for the Promotion of Science (JSPS), Japan Science and Technology Agency (JST), the Toray Science Foundation, NAOJ, Kavli IPMU, KEK, ASIAA, and Princeton University. 

This paper makes use of software developed for the Large Synoptic Survey Telescope. We thank the LSST Project for making their code available as free software at  http://dm.lsst.org

The Pan-STARRS1 Surveys (PS1) have been made possible through contributions of the Institute for Astronomy, the University of Hawaii, the Pan-STARRS Project Office, the Max-Planck Society and its participating institutes, the Max Planck Institute for Astronomy, Heidelberg and the Max Planck Institute for Extraterrestrial Physics, Garching, The Johns Hopkins University, Durham University, the University of Edinburgh, Queen’s University Belfast, the Harvard-Smithsonian Center for Astrophysics, the Las Cumbres Observatory Global Telescope Network Incorporated, the National Central University of Taiwan, the Space Telescope Science Institute, the National Aeronautics and Space Administration under Grant No. NNX08AR22G issued through the Planetary Science Division of the NASA Science Mission Directorate, the National Science Foundation under Grant No. AST-1238877, the University of Maryland, and Eotvos Lorand University (ELTE) and the Los Alamos National Laboratory.


\begin{thebibliography}{}
\bibitem[Abazajian et al.(2004)]{Abazajian04} Abazajian, K., Adelman-McCarthy, J.~K., Ag{\"u}eros, M.~A., et al.\ 2004, \aj, 128, 502 
\bibitem[Adams et al.(2015)]{Adams15} Adams, S.~M., Martini, P., Croxall, K.~V., Overzier, R.~A., \& Silverman, J.~D.\ 2015, \mnras, 448, 1335 
\bibitem[Adelberger \& Steidel(2005)]{Adelberger05} Adelberger, K.~L., \& Steidel, C.~C.\ 2005, \apj, 630, 50
\bibitem[Aihara et al.(2017)]{Aihara17a} Aihara, H., Armstrong, R., Bickerton, S., et al. submitted to PASJ (arXiv:1702.08449) 
\bibitem[Aihara et al.(2017)]{Aihara17b} Aihara, H., Arimoto, N., Armstrong, R., et al. to be submitted to PASJ  
\bibitem[Axelrod et al.(2010)]{Axelrod10} Axelrod, T., Kantor, J., Lupton, R.~H., \& Pierfederici, F.\ 2010, \procspie, 7740, 774015 
\bibitem[Angulo et al.(2012)]{Angulo12} Angulo, R.~E., Springel, V., White, S.~D.~M., et al.\ 2012, \mnras, 425, 2722 
\bibitem[Baldry et al.(2004)]{Baldry04} Baldry, I.~K., Glazebrook, K., Brinkmann, J., et al.\ 2004, \apj, 600, 681
\bibitem[Ba{\~n}ados et al.(2013)]{Banados13} Ba{\~n}ados, E., Venemans, B., Walter, F., et al.\ 2013, \apj, 773, 178 
\bibitem[Becker et al.(1995)]{Becker95} Becker, R.~H., White, R.~L., \& Helfand, D.~J.\ 1995, \apj, 450, 559 
\bibitem[Behroozi et al.(2013)]{Behroozi13} Behroozi, P.~S., Wechsler, R.~H., \& Conroy, C.\ 2013, \apj, 770, 57 
\bibitem[B{\"o}hringer et al.(2001)]{Boh01} B{\"o}hringer, H., Schuecker, P., Guzzo, L., et al.\ 2001, \aap, 369, 826 
\bibitem[Bond et al.(1991)]{Bond91} Bond, J.~R., Cole, S., Efstathiou, G., \& Kaiser, N.\ 1991, \apj, 379, 440 
\bibitem[Bosch et al.(2017)]{Bosch17} Bosch, J. et al. to be submitted to PASJ  
\bibitem[Bower(1991)]{Bower91} Bower, R.~G.\ 1991, \mnras, 248, 332 
\bibitem[Bruns et al.(2012)]{Bruns12} Bruns, L.~R., Wyithe, J.~S.~B., Bland-Hawthorn, J., \& Dijkstra, M.\ 2012, \mnras, 421, 2543 
\bibitem[Cai et al.(2016)]{Cai16} Cai, Z., Fan, X., Peirani, S., et al.\ 2016, \apj, 833, 135 
\bibitem[Chiang et al.(2013)]{Chiang13} Chiang, Y.-K., Overzier, R., \& Gebhardt, K.\ 2013, \apj, 779, 127 
\bibitem[Coil et al.(2007)]{Coil07} Coil, A.~L., Hennawi, J.~F., Newman, J.~A., Cooper, M.~C., \& Davis, M.\ 2007, \apj, 654, 115  
\bibitem[Cole et al.(2000)]{Cole00} Cole, S., Lacey, C.~G., Baugh, C.~M., \& Frenk, C.~S.\ 2000, \mnras, 319, 168 
\bibitem[Cooke et al.(2014)]{Cooke14} Cooke, E.~A., Hatch, N.~A., Muldrew, S.~I., Rigby, E.~E., \& Kurk, J.~D.\ 2014, \mnras, 440, 3262 
\bibitem[Coatman et al.(2017)]{Coatman17} Coatman, L., Hewett, P.~C., Banerji, M., et al.\ 2017, \mnras, 465, 2120 
\bibitem[Dawson et al.(2013)]{Dawson13} Dawson, K.~S., Schlegel, D.~J., Ahn, C.~P., et al.\ 2013, \aj, 145, 10 
\bibitem[Dekel \& Birnboim(2006)]{Dekel06} Dekel, A., \& Birnboim, Y.\ 2006, \mnras, 368, 2 
\bibitem[Donoso et al.(2010)]{Donoso10} Donoso, E., Li, C., Kauffmann, G., Best, P.~N., \& Heckman, T.~M.\ 2010, \mnras, 407, 1078 
\bibitem[Dressler(1980)]{Dressler80} Dressler, A.\ 1980, \apj, 236, 351
\bibitem[Eftekharzadeh et al.(2015)]{Eftek15} Eftekharzadeh, S., Myers, A.~D., White, M., et al.\ 2015, \mnras, 453, 2779 
\bibitem[Enoki et al.(2003)]{Enoki03} Enoki, M., Nagashima, M., \& Gouda, N.\ 2003, \pasj, 55, 133 
\bibitem[Enoki et al.(2014)]{Enoki14} Enoki, M., Ishiyama, T., Kobayashi, M.~A.~R., \& Nagashima, M.\ 2014, \apj, 794, 69 
\bibitem[Fan et al.(2006)]{Fan06} Fan, X., Strauss, M.~A., Becker, R.~H., et al.\ 2006, \aj, 132, 117 
\bibitem[Fanidakis et al.(2012)]{Fanidakis12} Fanidakis, N., Baugh, C.~M., Benson, A.~J., et al.\ 2012, \mnras, 419, 2797 
\bibitem[Fanidakis et al.(2013)]{Fanidakis13} Fanidakis, N., Macci{\`o}, A.~V., Baugh, C.~M., Lacey, C.~G., \& Frenk, C.~S.\ 2013, \mnras, 436, 315
\bibitem[Hatch et al.(2014)]{Hatch14} Hatch, N.~A., Wylezalek, D., Kurk, J.~D., et al.\ 2014, \mnras, 445, 280 
\bibitem[He et al.(2017)]{He17} He, W., Akiyama, M., Enoki, M. et al. to be submitted to PASJ  
\bibitem[Henriques et al.(2012)]{Henriques12} Henriques, B.~M.~B., White, S.~D.~M., Lemson, G., et al.\ 2012, \mnras, 421, 2904 
\bibitem[Hopkins et al.(2008)]{Hopkins08} Hopkins, P.~F., Hernquist, L., Cox, T.~J., \& Kere{\v s}, D.\ 2008, \apjs, 175, 356-389 
\bibitem[Husband et al.(2013)]{Husband13} Husband, K., Bremer, M.~N., Stanway, E.~R., et al.\ 2013, \mnras, 432, 2869
\bibitem[Ichikawa \& Inayoshi(2017)]{Ichikawa17} Ichikawa, K., \& Inayoshi, K.\ 2017, \apjl, 840, L9 
\bibitem[Ishikawa et al.(2015)]{Ishikawa05} Ishikawa, S., Kashikawa, N., Toshikawa, J., \& Onoue, M.\ 2015, \mnras, 454, 205 
\bibitem[Ivezic et al.(2008)]{Ivezic08} Ivezic, Z., Axelrod, T., Brandt, W.~N., et al.\ 2008, Serbian Astronomical Journal, 176, 1 
\bibitem[Magnier et al.(2013)]{Magnier13} Magnier, E.~A., Schlafly, E., Finkbeiner, D., et al.\ 2013, \apjs, 205, 20 
\bibitem[Mazzucchelli et al.(2017)]{Mazzucchelli17} Mazzucchelli, C., Ba{\~n}ados, E., Decarli, R., et al.\ 2017, \apj, 834, 83 
\bibitem[Juri{\'c} et al.(2015)]{Juric15} Juri{\'c}, M., Kantor, J., Lim, K., et al.\ 2015, arXiv:1512.07914 
\bibitem[Kashikawa et al.(2007)]{Kashikawa07} Kashikawa, N., Kitayama, T., Doi, M., et al.\ 2007, \apj, 663, 765 
\bibitem[Kawanomoto et al.(2017)]{Kawanomoto17} Kawanomoto, S. et al. to be submitted to PASJ  
\bibitem[Kere{\v s} et al.(2005)]{Keres05} Kere{\v s}, D., Katz, N., Weinberg, D.~H., \& Dav{\'e}, R.\ 2005, \mnras, 363, 2 
\bibitem[Kikuta et al.(2017)]{Kikuta17} Kikuta, S., Imanishi, M., Matsuoka, Y., et al.\ 2017, \apj, 841, 128 
\bibitem[Kim et al.(2009)]{Kim09} Kim, S., Stiavelli, M., Trenti, M., et al.\ 2009, \apj, 695, 809
\bibitem[Koyama et al.(2013)]{Koyama13} Koyama, Y., Kodama, T., Tadaki, K.-i., et al.\ 2013, \mnras, 428, 1551 
\bibitem[Lacey \& Cole(1993)]{Lacey93} Lacey, C., \& Cole, S.\ 1993, \mnras, 262, 627 
\bibitem[Lacey et al.(2015)]{Lacey15} Lacey, C.~G., Baugh, C.~M., Frenk, C.~S., et al.\ 2015, arXiv:1509.08473
\bibitem[Lin et al.(2008)]{Lin08} Lin, L., Patton, D.~R., Koo, D.~C., et al.\ 2008, \apj, 681, 232-243 
\bibitem[Magliocchetti et al.(2002)]{Mag02} Magliocchetti, M., Maddox, S.~J., Jackson, C.~A., et al.\ 2002, \mnras, 333, 100 
\bibitem[Magorrian et al.(1998)]{Magorrian98} Magorrian, J., Tremaine, S., Richstone, D., et al.\ 1998, \aj, 115, 2285 
\bibitem[Malkan \& Sargent(1982)]{Malkan82} Malkan, M.~A., \& Sargent, W.~L.~W.\ 1982, \apj, 254, 22 
\bibitem[Marconi \& Hunt(2003)]{Marconi03} Marconi, A., \& Hunt, L.~K.\ 2003, \apjl, 589, L21 
\bibitem[Martini(2004)]{Martini04} Martini, P.\ 2004, Coevolution of Black Holes and Galaxies, 169 
\bibitem[Mauch \& Sadler(2007)]{Mauch07} Mauch, T., \& Sadler, E.~M.\ 2007, \mnras, 375, 931  
\bibitem[Miyazaki et al.(2012)]{Miyazaki12} Miyazaki, S., Komiyama, Y., Nakaya, H., et al.\ 2012, \procspie, 8446, 84460Z 
\bibitem[Morselli et al.(2014)]{Morselli14} Morselli, L., Mignoli, M., Gilli, R., et al.\ 2014, \aap, 568, A1  
\bibitem[Mortlock et al.(2011)]{Mortlock11} Mortlock, D.~J., Warren, S.~J., Venemans, B.~P., et al.\ 2011, \nat, 474, 616 
\bibitem[Nagashima et al.(2005)]{Nagashima05} Nagashima, M., Yahagi, H., Enoki, M., Yoshii, Y., \& Gouda, N.\ 2005, \apj, 634, 26 
\bibitem[Ocvirk et al.(2008)]{Ocvirk08} Ocvirk, P., Pichon, C., \& Teyssier, R.\ 2008, \mnras, 390, 1326 
\bibitem[Ono et al.(2017)]{Ono17} Ono, Y. et al. to be submitted to PASJ  
\bibitem[Onoue et al.(2017)]{Onoue17} Onoue, M., Kashikawa, N., Uchiyama, H. et al. to be submitted to PASJ  
\bibitem[Oogi et al.(2016)]{Oogi16} Oogi, T., Enoki, M., Ishiyama, T., et al.\ 2016, \mnras, 456, L30 
\bibitem[Orsi et al.(2016)]{Orsi16} Orsi, {\'A}.~A., Fanidakis, N., Lacey, C.~G., \& Baugh, C.~M.\ 2016, \mnras, 456, 3827 
\bibitem[Overzier(2016)]{Overzier2016} Overzier, R.~A.\ 2016, \aapr, 24, 14 
\bibitem[Park et al.(2013)]{Park2013} Park, D., Woo, J.-H., Denney, K.~D., \& Shin, J.\ 2013, \apj, 770, 87 
\bibitem[P{\^a}ris et al.(2017)]{Paris17} P{\^a}ris, I., Petitjean, P., Ross, N.~P., et al.\ 2017, \aap, 597, A79
\bibitem[Press \& Schechter(1974)]{Press74} Press, W.~H., \& Schechter, P.\ 1974, \apj, 187, 425 
\bibitem[Ross et al.(2012)]{Ross12} Ross, N.~P., Myers, A.~D., Sheldon, E.~S., et al.\ 2012, \apjs, 199, 3 
\bibitem[Rozo et al.(2009)]{Rozo09} Rozo, E., Rykoff, E.~S., Evrard, A., et al.\ 2009, \apj, 699, 768 
\bibitem[Schlafly et al.(2012)]{Schlafly12} Schlafly, E.~F., Finkbeiner, D.~P., Juri{\'c}, M., et al.\ 2012, \apj, 756, 158 
\bibitem[Shen et al.(2007)]{Shen07} Shen, Y., Strauss, M.~A., Oguri, M., et al.\ 2007, \aj, 133, 2222 
\bibitem[Shen et al.(2008)]{Shen08} Shen, Y., Greene, J.~E., Strauss, M.~A., Richards, G.~T., \& Schneider, D.~P.\ 2008, \apj, 680, 169-190 
\bibitem[Shen et al.(2013)]{Shen13} Shen, Y., McBride, C.~K., White, M., et al.\ 2013, \apj, 778, 98 
\bibitem[Shirakata et al.(2015)]{Shirakata15} Shirakata, H., Okamoto, T., Enoki, M., et al.\ 2015, \mnras, 450, L6 
\bibitem[Simpson(2005)]{Simpson05} Simpson, C.\ 2005, \mnras, 360, 565  
\bibitem[Springel et al.(2005)]{Springel05} Springel, V., White, S.~D.~M., Jenkins, A., et al.\ 2005, \nat, 435, 629 
\bibitem[Steidel et al.(2005)]{Steidel05} Steidel, C.~C., Adelberger, K.~L., Shapley, A.~E., et al.\ 2005, \apj, 626, 44 
\bibitem[Tonry et al.(2012)]{Tonry12} Tonry, J.~L., Stubbs, C.~W., Lykke, K.~R., et al.\ 2012, \apj, 750, 99 
\bibitem[Toshikawa et al.(2016)]{Toshikawa16} Toshikawa, J., Kashikawa, N., Overzier, R., et al.\ 2016, \apj, 826, 114 
\bibitem[Toshikawa et al.(2017)]{Toshikawa17} Toshikawa, J., et al. to be submitted to PASJ
\bibitem[Trainor \& Steidel(2012)]{Trainor12} Trainor, R.~F., \& Steidel, C.~C.\ 2012, \apj, 752, 39 
\bibitem[Utsumi et al.(2010)]{Utsumi10} Utsumi, Y., Goto, T., Kashikawa, N., et al.\ 2010, \apj, 721, 1680 
\bibitem[van der Burg et al.(2010)]{vanderBurg10} van der Burg, R.~F.~J., Hildebrandt, H., \& Erben, T.\ 2010, \aap, 523, A74 
\bibitem[Venemans et al.(2007)]{Venemans07} Venemans, B.~P., R{\"o}ttgering, H.~J.~A., Miley, G.~K., et al.\ 2007, \aap, 461, 823 
\bibitem[Vestergaard(2002)]{Vestergaard2002} Vestergaard, M.\ 2002, \apj, 571, 733 
\bibitem[White et al.(2012)]{White12} White, M., Myers, A.~D., Ross, N.~P., et al.\ 2012, \mnras, 424, 933 

\end{thebibliography}
\end{document}